\documentclass[aps,prl]{revtex4}  %% REVTeX 4.0
\usepackage{dcolumn}% Align table columns on decimal point
\usepackage{bm}% bold math

\begin{document}

\title{Generation of entangled 3D localized quantum wave-packets via optical parametric amplification}% Forfoce line breaks with \\
\author{Claudio Conti}
\email{c.conti@ele.uniroma3.it}
\affiliation{%
Nonlinear Optics and Optoelectronics Labs (N{\it OO}EL)\\
National Institute for the Physics of the Matter (INFM) - Roma Tre
}%

\date{\today}% It is always \today, today,

\begin{abstract}
In the framework of the paraxial and of the slowly varying envelope approximations, 
with reference to a normally dispersive medium or to vacuum, 
the electromagnetic field is given 
 as a continuous quantum superposition of non-dispersive and non-diffracting 
wave-packets (namely X-waves).  
Entangled states as pairs of elementary excitations traveling at (approximately) the same 
velocity are found in optical parametric amplification.
\end{abstract}
%\ocis{000.0000, 999.9999.}% REPLACE WITH CORRECT OCIS CODES FOR YOUR ARTICLE
                          % NOTE: \ocis{} IS ALIASED TO \pacs{} BUT MUST
                          % FORMAT THE TERMS CORRECTLY FOR EACH JOURNAL

\maketitle
\section{Introduction}
The quantum description of diffracting non-monochromatic beams has
been the subject of intense research 
(see e.g. references \cite{Chiao91,Huttner92,Kennedy88,Drummond99,Dung98}).
Various approaches have been proposed in order to provide
a consistent quantum picture of nonlinear optical processes and
related phenomena, like entanglement and photon localization.
Several important issues have been discussed in the 
literature, as equal-time or equal-space commutation rules, or paraxial approximations,
and different ``schemes'' have been formulated.
\cite{Kozlov01,Haus90,Deutsch91,Crosignani95,Hagelstein96} 
Most of them are based on the second quantization of the relevant classical equations,
or on a three dimensional plane wave expansion,
however their employment in specific problems is far from being trivial.
In this Letter I describe a new formulation for 3D optical field quantization, 
and discuss its application to optical parametric amplification. 

All the machinery is based on the paraxial and slowly varying envelope 
approximation (SVEA), with reference to vacuum and to 
a normally dispersive medium. However a general theory can
be envisaged. The basic element is recognizing that a traveling 
pulsed beam
can be expressed as a continuous set of harmonic oscillators, each
weighted by a rigidly moving 3D wave-packet. It corresponds to a non-dispersive
non-diffracting solution of the relevant classical wave equation, with velocity $v$. 
This is the so-called X-wave transform, 
or X-wave expansion,  as outlined
below. \cite{Lu00,Salo01b}  Then the quantization comes naturally, following standard approach, 
using continuous-mode operators. \cite{DiracBook,Blow90}

X-waves are 3D invariant pulsed-beams traveling without diffraction and dispersion, also known to as 
``progressive undistorted waves''  (see e.g. the recent reviews \cite{Saari01,Recami03}
and references therein). Their appearance in nonlinear optical processes
has been experimentally investigated in \cite{DiTrapani03}.

For free propagation, it turns out that the quasi-monochromatic beam, 
with carrier angular frequency $\omega$, behaves like a quantum gas of
particles which, in vacuo, have mass $m$ satisfying the relation
$mc^2=\hbar\omega$. The rigidly moving ``eigenmodes'' (the X-waves), while not
having a finite norm (like plane waves), do have a strong degree of 
localization, thus addressing the issue of spatio-temporal 
photon confinement (see e.g. \cite{Chan02} and references therein).
In a simple way, this approach provides the quantum description of the 
3D evolution of a classical, paraxial and slowly varying, beam.
Furthermore classical infinite-energy X-waves, here described as coherent states, 
do have finite expectation values for energy in the quantized picture.

\section{Derivation of the paraxial equations}
In this article a new quantum-mechanical 
description of pulsed light beams is introduced. In order 
to avoid as much as possible unnecessary complexities and involved formalism, 
paraxial slowly varying beams are considered. 
Given the fact that progressive undistorted waves are well-studied solutions 
of the wave equation, a much more general theory can be envisaged.
Typical experiments in nonlinear optics are described in the 
framework of the considered approximations and, for similar reasons, the treatment is limited to circularly symmetric beams.

The basic model can be obtained by the scalar 3D wave equation for the electromagnetic 
field $E$ in a medium with refractive index $n$:
\begin{equation}
\label{Maxwell}
\nabla^2E-\frac{n^2}{c^2}\frac{\partial^2 E}{\partial t^2}=0\text{.}
\end{equation}
By letting $E=Re[A\exp(-i\omega t+ik z)]/\sqrt{\epsilon_0 n^2/2}$, 
with $k=\omega n/c$, in the paraxial approximation
the usual Foch-Leontovich equation can be derived:
\begin{equation}
\label{FochLeontovich}
2ik\frac{\partial A}{\partial z}+
2i\frac{\omega}{c^2}\frac{\partial A}{\partial t}+
\nabla_\perp^2 A-\frac{1}{c^2}\frac{\partial^2 A}{\partial t^2}=0\text{.}
\end{equation}
Equations like (\ref{FochLeontovich}) are typically adopted in nonlinear optics, and
interpreted as a propagation problem along $z$. If $A$, as a function of $t$, is slowly varying, 
they can be equivalently cast as a time-evolution problem;
the latter is the formulation adopted here in order to be as much as possible consistent
with standard quantum mechanics.
With this aim, introducing $\zeta=z-ct$ and $\tau=t$, Eq. (\ref{Maxwell}) is rewritten as 
\begin{equation}
\label{FochLeontovich2}
2i\frac{\omega}{c^2}\frac{\partial A}{\partial \tau}+
\nabla_\perp^2 A-\frac{\partial^2 A}{\partial \zeta^2}=
\frac{1}{c^2}\frac{\partial^2 A}{\partial \tau^2}
-\frac{2}{c}\frac{\partial^2 A}{\partial \tau\partial \zeta}\text{.}
\end{equation}
If the SVEA is applicable, the right-hand side
of the previous equation is negligible and, coming back to the original variables
$(z,t)$, Eq. (\ref{FochLeontovich2}) can be written as
\begin{equation}
\label{main1}
i\frac{\partial A}{\partial t}+i \omega' \frac{\partial A}{\partial z}
-\frac{\omega''}{2}\frac{\partial A}{\partial z^2}
+\frac{\omega'}{2 k}\nabla^2_{\perp}A=0\text{,}
\end{equation}
while being $\omega'=c$ and $\omega''=c^2/\omega$.
Eq (\ref{main1}) can be treated as an evolution problem with respect to time $t$.

The very same procedure can be repeated in a dispersive medium, as done in Ref. \onlinecite{Kennedy88},
while retaining terms up to the second order dispersion.  The model (\ref{main1}) still holds
with $\omega'$ and $\omega''$ the derivatives of the dispersion relation $\omega=\omega(k)$. 
Hence,
since it can describe vacuum as well as a dispersive material,
\footnote{For uniformity, I will limit the treatment to 
a normally dispersive material ($\omega''>0$).} and it is typically
used with reference to experiments with pulsed-diffracting beams, it will be 
the starting point for the quantization procedure. 
Without loss of generality the envelope field $A$ has been normalized so that 
the time averaged electromagnetic energy in the whole 3D space is 
\begin{equation}
\label{energy0}
\mathcal{E}=\int\int\int |A|^2 dxdydz\text{,}
\end{equation}
i.e. any type of box-normalization is not necessary. 
\section{The X-wave transform and X-wave expansion}
Assuming radial symmetry hereafter ($r\equiv\sqrt{x^2+y^2}$), the general solution for the initial value problem is expressed by the Fourier-Bessel 
integral
\begin{equation}
\label{planewavesolution}
A=\int_{-\infty}^{\infty}\int_{0}^{\infty}k_\perp J_0(k_\perp r) S(k_\perp,k_z)
e^{ik_z Z-i\Omega t}dk_\perp dk_z\text{,}
\end{equation}
with $Z=z-\omega't$.
Eq. (\ref{planewavesolution})
 furnishes the field at instant $t$, given its spectrum $S$ at $t=0$, with transversal and longitudinal wave-numbers $k_\perp$ and $k_z$, and $\Omega=-\omega'' k_z^2/2+\omega' k_\perp^2/2k$.
With the following change of variables ($\alpha>0$) 
\begin{equation}
\label{newvariables} 
\begin{array}{l}
k_\perp=\alpha\sqrt{\omega''k/\omega'}\\
k_z=\alpha-v/\omega''\text{,}
\end{array}
\end{equation} 
I have, after some manipulations,
\begin{equation}
\label{Xtransform}
A=\int_{-\infty}^{\infty} e^{-i\frac{v^2}{2\omega''}t}\Psi_v(r,Z-vt)dv\text{,}
\end{equation}  
while being
\begin{equation}
\label{Psi}
\Psi_v(r,Z-vt)=\int_0^{\infty} X(\alpha,v)J_0(\sqrt{\frac{\omega''k}{\omega'}}\alpha r)
e^{i(\alpha-\frac{v}{\omega''})(Z-vt)}d\alpha\text{.}
\end{equation}
Hence $A$ can be expressed as a superposition of self-invariant beams (i.e. waves traveling along the $z-$direction
without dispersion and diffraction) with different velocities. 
The quantity $X(\alpha,v)=\frac{k\alpha}{\omega'}S(\alpha 
\sqrt{\frac{\omega'' k}{\omega'}},\alpha-\frac{v}{\omega''})$ corresponds 
to the so-called {\it X-wave transform}.\cite{Lu00}
 X-waves are defined as the propagation 
invariant solutions of the wave equation (\ref{main1}), for more details see
Ref. \onlinecite{Conti03} and references therein.

$X(\alpha,v)$ can be expanded 
by Laguerre functions \cite{Salo01b} $L_p^{(1)}(2x)e^{-x}$ with respect to $\alpha$:
\begin{equation}
\label{fs}
\begin{array}{l}
X(\alpha,v)=\Sigma_{p=0}^{\infty}C_p(v) f_p(\alpha)\\
\\
f_p(\alpha)=\sqrt{\frac{k}{\pi^2\omega'(p+1)}}(\alpha \Delta) L_p^{(1)}(2\alpha\Delta)e^{-\alpha\Delta}\text{,}
\end{array}
\end{equation}
where $\Delta$ is a reference length that is related to the spatial extension of the
beam (the specific form of $f_p$ is chosen for later convenience). $A$ is written as a superposition of basis orthogonal X-waves (given by 
(\ref{Psi}) with $X(\alpha,v)=f_p(\alpha)$):
\begin{equation}
\label{classicalfield}
A=\Sigma_{p=0}^{\infty}\int_{-\infty}^{\infty}C_p(v)e^{-i\frac{v^2}{2\omega''}t}
\psi_p^v(r,Z-vt)dv\text{.}
\end{equation}
The following relation holds by construction
\begin{equation}
\label{ortX}
<\psi_q^u(r,Z-ut)|\psi_p^v(r,Z-vt)>=\delta_{pq}\delta(u-v)\text{,}
\end{equation}
where $<f|g>$ denotes the integral over $x,y,z$ of $f^*\,g$.
Eq. (\ref{ortX}) reveals that the X-waves have infinite norm, 
like the plane waves typically adopted for field quantization.

\section{Quantization}
The classical energy of the pulsed beam, Eq. (\ref{energy0}), is given by
\begin{equation}
\label{energy}
\mathcal{E}=\Sigma_{p=0}^{\infty}\int_{-\infty}^{\infty}|C_p(v)|^2dv\text{,}
\end{equation}
while the field in (\ref{classicalfield}) can be rewritten as
\begin{equation}
\label{classicalfield2}
A=\Sigma_{p=0}^{\infty}\int_{-\infty}^{\infty}C_p(v,t)
\psi_p^v(r,Z-vt)dv\text{,}
\end{equation}
being 
\begin{equation}
\label{complexoscillators}
\frac{dC_p}{dt}(v,t)=-i\omega_p(v) C_p(v,t)
\end{equation}
with $\omega_p=v^2/2\omega''$.
Hence $A$ is the integral sum of harmonic oscillators, with complex amplitudes
$C_p$, each associated to a traveling invariant beam (corresponding to
the usual cavity mode, with the difference that it is rigidly moving).
Introducing the  real-valued 
``position'' $Q_p(v)$ and ``momentum'' $P_p(v)$, by letting 
\begin{equation}
\label{position_momentum}
C_p(v,t)=\frac{1}{\sqrt{2}}[\omega_p(v)Q_p(v,t)+i P_p(v,t)]\text{,}
\end{equation}
$\mathcal{E}$ is written as the sum of the time-independent classical energies of each oscillator
\begin{equation}
\label{energy2}
\mathcal{E}=\Sigma_{p=0}^{\infty}\int_{-\infty}^{\infty}
\frac{1}{2}[P_p(v,t)^2+\omega_p(v)^2 Q_p(v,t)^2]
dv\text{,}
\end{equation}
$\omega_p(v)$ is the resonant frequency of each ``mode''. 

Following standard approach to quantization, \cite{ItzyksonBook,LoudonBook} 
it is natural to write the Hamiltonian like
(omitting extrema for the integral and the $\Sigma$ symbol)
\begin{equation}
\label{Hamiltonian}
H=\Sigma_p \int \hbar \omega_p(v) a_p^{\dagger}(v) a_p(v)dv=
\Sigma_p \int \frac{m v^2}{2} a_p^{\dagger}(v) a_p(v)dv
\text{,}
\end{equation}
with $[a_q(u),a_p^{\dagger}(v)]=\delta_{pq}\delta(u-v)$. 
Note that the zero-point energy has been omitted in (\ref{Hamiltonian}),
as a standard renormalization in quantum field theory. \cite{ItzyksonBook}

Some comments are needed at this point.
First of all I observe that (\ref{Hamiltonian}) corresponds to 
a set of freely moving particles, each contributing to 
the energy by the kinetic term, with mass defined by
\begin{equation}
\label{mass1}
m=\frac{\hbar}{\omega''}\text{.}
\end{equation}
Thus the freely propagating pulsed beam is represented
like a quantum fluid, or gas.\cite{Landau} The eigenstates are given by 
$|l,n,v>=[a_l^{\dagger}(v)]^n|0>$, denoting $n$ particles with velocity $v$ in 
the traveling ``mode''  (or fundamental X-wave) $\psi_l^v$, and energy $n mv^2/2$.
Then, it is to note that, differently from standard heuristic approaches to field
quantization, the Hamiltonian directly corresponds to the energy of the beam, 
and not to the second quantized form, i.e. to the 
Hamiltonian functional giving Eq. (\ref{main1}). 

The optical field amplitude operator is given by
\begin{equation}
\label{fieldoperator}
A=\Sigma_p\int e^{-\frac{i}{\hbar}\frac{mv^2}{2}t}
\,\psi_p^v[z-(\omega'+v)t,r]\sqrt{\hbar\omega_p(v)}a_p(v)dv\text{.}
\end{equation}
The mean value of $A$ and of the energy density,
in the state with one elementary excitation on the $l-$fundamental X-wave at velocity $v$, are: 
 $<l,1,v|A|l,1,v>=0$ and $<l,1,v|A^{\dagger}A|l,1,v>=\hbar\omega_l(v)|\psi_l^v(Z-vt,r)|^2$.

Coherent states can we constructed as well.
For example, taking  (with obvious notation)
\begin{equation}
|\alpha_l(v)>=e^{-\frac{|\alpha_l(v)|^2}{2}}
\Sigma_{n=0}^\infty\frac{\alpha_l(v)^n}{\sqrt{n!}}|l,n,v>\text{,}
\end{equation}
with $\alpha_l(v)$ an arbitrary complex number,
I have a state that has an expectation value for the optical field profile given by the
fundamental X-wave $\psi_l^v(Z-vt,r)$ and 
corresponds to a classical X-wave.
\footnote{A generic optical field will be given by a superposition of these states.}
However there is a remarkable difference:
the expectation value 
of the energy is $\hbar\omega_l(v)|\alpha_l(v)|^2$.
While this state has infinite energy in a classical context,
{\it in a quantum mechanical picture X-waves do have finite energy}.

On the other hand it is well known, from quantum field theory, that $|l,n,v>$ is not 
normalizable, and  $a_l^\dagger(v)$ must be intended as operator valued distributions. \cite{ItzyksonBook}
This is due to the fact that a state with a definite $v$ is an idealization, like an elementary particle 
with a definite momentum $k$. Any experimental apparatus implies a spread, and an indeterminacy in the velocity
(corresponding to a normalized state obtained as a continuous superposition of $|l,n,v>$). 

By particularizing Eq. (\ref{mass1}) to the 
case of vacuum, using the previously mentioned relations, it is found
\begin{equation}
\label{einstein}
m\,c^2=\hbar\omega\text{,}
\end{equation}
which shows that the effective mass of the quantum gas particles is determined 
by the carrier frequency of the beam via the Einstein relation.

It is also notable that, in vacuo ($\omega'=c$), superluminal components are encompassed in
Eq. (\ref{fieldoperator}). But this is not surprising. Indeed
the superluminal propagation of X-waves does not lead to paradoxes, since they
are stationary solutions (i.e. they fill all the space, like
plane waves) of the wave equations (see also \cite{Ciattoni03,Recami03}). 
This can be shown to be perfectly consistent with the principles of 
standard special relativity, 
in the framework of the covariant Maxwell equations.\cite{CiattoniPrivate} Furthermore 
it is well known that superluminal propagation has a not zero probability, 
and hence it is predicted, in standard quantum electrodynamics. \cite{Feynman} 
Single-photon superluminal effects have been experimentally investigated in Ref. \onlinecite{Steinberg93}. 
\footnote{Obviously the paraxial-SVEA model is not Lorentz invariant,
the relevant generalization will be the subject of forthcoming papers. }
Before considering an application of this formalism I want to underline that 
a completely different wavelet quantization of relativistic 
fields has been previously developed by Kaiser.\cite{Kaiser03} 

\section{Optical Parametric Amplification}
I consider the phase-matched co-directional optical parametric
amplification (OPA) \cite{Hong85} of two frequencies, $\omega_1$ and $\omega_2$, attainable in 
quadratic, or cubic, nonlinear media. For simplicity the pump beam is treated 
as a constant term in the Hamiltonian. A quantized field is associated to $\omega_1$ and $\omega_2$, with parameters 
$\omega'_{1,2}$, $k_{1,2}=\omega_{1,2}n_{1,2}/c$. $\omega''$ is taken 
equal for both of them, and $\omega_1'\neq\omega_2'$. $A_{1,2}$ denote the corresponding optical quantum fields:
\begin{equation}
\begin{array}{l}
A_1=\Sigma_p \int e^{-i\frac{v^2\,t}{2\omega''}}\psi_{p,1}^{v}[z-(\omega_1'+v)t,r]
\sqrt{\hbar\omega_p(v)}a_{p}(v)dv\\
A_2=\Sigma_p \int e^{-i\frac{v^2\,t}{2\omega''}}\psi_{p,2}^{v}[z-(\omega_2'+v)t,r]
\sqrt{\hbar\omega_p(v)}b_{p}(v)dv\text{.}
\end{array}
\end{equation}
The basis X-waves are denoted by $\psi_{p,j}^{v}$ in order to indicate 
that they have to be calculated after Eq. (\ref{fs}) with parameters ($\omega_j'$ and $k_j$) of $\omega_j$; the 
corresponding spectra are denoted by $f_{p,j}(\alpha)$ ($j=1,2$).

The total Hamiltonian is given by the sum of the free terms and the interaction Hamiltonian $H_I$:
\begin{equation}
H=\Sigma_p\int \hbar\omega_p(v)(a_p^\dagger a_p+b_p^\dagger b_p)dv+H_I\text{.}
\end{equation}
$H_I$ can be determined by the classical counterpart given by
$\mathcal{E}_I=\chi<A_1|A_2^*>+c.c.$ ($c.c.$ denotes complex conjugation).
After some (lengthy but straightforward)
 manipulations ($h.c.$ is hermitian conjugate):
\begin{equation}
\label{interaction}
H_I=\hbar\Sigma_{pq}\int\int du dv
\chi_{pq}(u+v)\sqrt{\omega_p(v)\omega_q(u)}
e^{i F(u,v)t} a_p^{\dagger}(u)b_q^{\dagger}(v)+\text{h.c.}
\end{equation}
where
$F(u,v)=(u^2+v^2)/2\omega''+
(u-v+\omega_1'-\omega_2')(v-\rho u)/\omega''$, with $\rho=(k_1\omega_2'/k_2\omega_1')^{1/2}$.
The interaction term $\chi_{pq}(\nu)$ is
\begin{equation}
\chi_{pq}(\nu)=
\frac{4\pi^2\chi}{\nu}\sqrt{\frac{\omega_1'\omega_2'}{k_1 k_2}} 
f_{p,1}[\frac{\nu}{(1+\rho)\omega''}]
f_{q,2}[\frac{\rho\nu}{(1+\rho)\omega''}]\theta(\nu)\text{,}
\end{equation}
with $\theta(\nu)$ the unit step function.

By applying standard time-dependent perturbation theory, assuming 
as the initial state $|0>$ at $t=0$, at first order it is
\begin{equation}
\label{pertstate}
\begin{array}{l}
|\psi^{(1)}(t)>=
\Sigma_{pq}\int\int
e^{iK(u,v)t} \chi_{pq}(u+v)
G(u,v,t) \sqrt{\omega_p(v)\omega_q(u)}a_p^{\dagger}(u)b_q^{\dagger}(v)dudv|0>\text{,}
\end{array}
\end{equation} 
with 
\begin{equation}
\label{Gdefinitions}
\begin{array}{l}
K(u,v)=\frac{(u-v+\omega_1'-\omega_2')(v-\rho u)}{2(1+\rho)\omega''}\\
g(u,v)=\frac{(u^2+v^2)}{\omega''}+\frac{(u-v+\omega_1'-\omega_2')(v-\rho u)}{(1+\rho)\omega''}\\
G(u,v,t)=\frac{2\sin[g(u,v)t/2]}{g(u,v)}\text{.}
\end{array}
\end{equation}

The state given in Eq. (\ref{pertstate}) is a continuous-variables 
entangled  superposition of 
particles at the two generated frequencies, traveling with different velocities.
Indeed the weight function is not separable with respect to $u$ and $v$
(see e.g. Ref. \onlinecite{Chan02} and references therein).

The transition probability to the two particles state 
$a_p^{\dagger}(u)b_q^{\dagger}(v)|0>$ is
given by 
\begin{equation}
\mathcal{P}_{pq}(t,u,v)=\omega_p(v)\omega_q(u)|\chi_{pq}(u+v)|^2
\frac{sin[g(u,v)t/2]^2}{[g(u,v)t/2]^2}t^2.
\end{equation}
As $t\rightarrow\infty$,  
$\mathcal{P}_{pq}$ tends to a Dirac-delta peaked at $g(u,v)=0$. 
Entangled particles, with velocities $u$ and $v$, 
are associated, for large propagation distances, to a point in the plane $(u,v)$, lying
on the parabola $g(u,v)=0$. 
\footnote{Related results, at a classical level, seem to be those of Ref. \onlinecite{Picozzi02},
where coherence along space-time trajectories in parametric wave mixing has been investigated.}
Of particular relevance is the low velocity region.

Proceeding as in quantum fluid theory, the small momenta approximation
can be applied.\cite{Landau} 
This means neglecting the quadratic terms in $u,v$ in $g(u,v)$,
which corresponds to small velocities, consistently with the SVEA.
In this case $g\cong(v-\rho u)(\omega_1'-\omega_2')/(1+\rho)\omega''$,
showing that, as $t\rightarrow\infty$,
\begin{equation}
\mathcal{P}_{pq}(t,u,v)\rightarrow
\omega_p(v)\omega_q(u)|\chi_{pq}(u+v)|^2
\frac{2\pi(1+\rho)\omega''t}{|\omega_1'-\omega_2'|}\,\delta(v-\rho u)\text{.}
\end{equation}
Thus, in the asymptotic state the two particles travel 
approximately at the same velocity ($\rho\cong1$
in practical cases). 

For large propagation distances, the quantum fluid is hence a superposition
of entangled pairs locked together.
This gives a clear picture of what entanglement is to be intended for. 
After the generation, the two excitations are associated to 
 X-shaped spatial distributions of energy that travel at the same $v$. 
Since the velocity determines the angular aperture (far from the origin)
 of the double cone which forms the X-wave,\cite{Ciattoni03} a three-dimensional region
where the two particles can be revealed simultaneously is determined. The duration
of the temporal profile fixes the useful time window.
In some sense, the paired elementary excitations are delocalized in an highly
localized 3D X-shaped rigidly moving region. 

\section{Conclusions}

Using X-waves as basis, a traveling wave quantization leads to the representation
of a pulsed beam as a quantum gas of free particles.
A structured, localized, spatio-temporal distribution of energy
is associated to each fundamental excitation. 
These elements furnish an alternative point of view regarding the concept of ``photon'',
and its localization, without introducing box-normalization or related ideas.

A 3D quantum propagation is reduced to a one dimensional evolution, thus enabling 
simple treatments of nonlinear optical processes. This is mainly a consequence of the 
paraxial approximation, which unavoidably fixes a preferential direction.
For this reason, this approach is well suited for describing optical experiments.
It is also susceptible of generalizations in Lorentz covariant theories.

During optical parametric amplification the particles composing the pulsed beam
become entangled in pairs moving, approximately, at the same velocity.
These excitations are delocalized into a 3D rigidly moving conical region,
where they can be revealed simultaneously.
The role of first order and of second order dispersion is clearly stated, while
including diffraction.
This provide a complete spatio-temporal picture of a nonlinear quantum process.

\section*{Aknowledgements}
I thank G. Assanto, A. Ciattoni, B. Crosignani, and S. Trillo for invaluable discussions.

%%Do not include separate BibTeX files; if BibTeX is used, paste the output here.
%\bibliographystyle{osa}
%\bibliography{longquantumX}

\end{document}